\documentclass{jfm}
\usepackage{graphicx,color}
\usepackage{epstopdf, epsfig}
\usepackage{appendix}

\def\NEW#1{\textcolor{black}{#1}}
\def\NNEW#1{\textcolor{black}{#1}}

\shorttitle{Suppression of inverse cascade in instability-driven 2-D turbulence}
\shortauthor{A. van Kan, B. Favier,  K. Julien and E. Knobloch}

\title{Spontaneous suppression of inverse energy cascade in instability-driven 2D turbulence}

\author{Adrian van Kan\aff{1}
  \corresp{\email{avankan@berkeley.edu}},
  Benjamin Favier\aff{2}, Keith Julien\aff{3}, Edgar Knobloch\aff{1}}

\affiliation{\aff{1}Department of Physics, University of California at Berkeley, Berkeley, California 94720, USA
\aff{2}Aix Marseille Univ., CNRS, Centrale Marseille, IRPHE, Marseille, France
\aff{3}Department of Applied Mathematics, University of Colorado, Boulder, CO 80309, USA}
\begin{document}

\maketitle

\begin{abstract}
Instabilities of fluid flows often generate turbulence. Using extensive direct numerical simulations, we study two-dimensional turbulence driven by a wavenumber-localised
instability superposed on stochastic forcing, in contrast to previous studies of state-independent forcing. As the contribution of the instability forcing, measured by a parameter $\gamma$, increases, the system undergoes two transitions. For $\gamma$ below a first threshold, a regular large-scale vortex condensate forms. Above this threshold, shielded vortices (SVs) emerge within the condensate. At a second, larger value of $\gamma$, the condensate breaks down, and a gas of weakly interacting vortices with broken symmetry spontaneously emerges, characterised by preponderance of vortices of one sign only and suppressed inverse energy cascade. The latter transition is shown to depend on the damping mechanism. The number density of SVs in the broken symmetry state slowly increases via a random nucleation process. Bistability is observed between the condensate and mixed SV-condensate states. Our findings provide new evidence for a strong dependence of two-dimensional turbulence phenomenology on the forcing.\end{abstract}


\section{Introduction}
Two-dimensional (2D) and quasi-2D flows arise in many systems, from \NEW{soap films \citep{vorobieff1999soap} to the Earth's atmosphere and oceans \citep{vallis2017atmospheric}}. 
Additional interest stems from active fluids, where suspended energy-consuming microswimmers can generate vortices and jets \citep{dombrowski2004self}. The basic phenomenology of 2D turbulence was developed by \cite{kraichnan1967inertial} who predicted that in such flows energy will be transferred from small to large scales, leading to an inverse energy cascade. This prediction was subsequently confirmed in direct numerical simulations (DNS) \citep{lilly1969numerical} 
and experiments \citep{sommeria1986experimental}. 
In finite domains, inverse cascades generate large-scale coherent structures, typically vortices or jets, called \textit{condensates} \NEW{\citep{smith1993bose}}. 

Beyond 2D turbulence, inverse cascades arise in highly anisotropic 3D flows in thin layers \citep{smith1996crossover}, 
rapidly rotating flows \citep{deusebio2014dimensional}, 
strongly stratified flows \citep{sozza2015dimensional}, among others. 
Inverse energy cascades in quasi-2D turbulence may also lead to a condensate if damping at large scales is small \citep{van2019condensates}. Condensates also arise in DNS of bacterial turbulence \citep{linkmann2019phase} and rapidly rotating convection \citep{rubio2014upscale}, 
 among others. Recent reviews of 2D and quasi-2D turbulence are given in \cite{boffetta2012two} and \cite{alexakis2018cascades}.  

The study of flows driven by a prescribed body force has a long history. Examples include time-independent forcing as in the Kolmogorov flow, 
 or random forcing with a prescribed energy injection rate. 
 Situations where the driving is prescribed independently of the flow configuration, as in these two examples, are attractive since they are often amenable to a detailed analysis. However, many real fluid flows are driven by instabilities, for instance of convective, shear or baroclinic type \citep{chandrasekhar2013hydrodynamic,salmon1980baroclinic}. Similarly, models of active fluid flows feature scale-dependent viscosities which can be negative at certain scales \citep{slomka2017geometry}, a fact consistent with the measured rheology of such flows \citep{lopez2015turning}. For instability-driven flows, the forcing explicitly depends on the velocity field and the injected power is proportional to the amplitude of the forcing-scale modes. \NEW{In contrast, the small-scale statistics of 3D turbulence with hybrid forcing are mostly forcing-independent \citep{lundgren2003linearly}.} 

Flows resulting from instabilities can differ drastically from Kraichnan's picture of the inverse cascade and condensation. For instance, active flows usually do not display an inverse cascade, but form mesoscale vortices \citep{wensink2012meso}.
Such coherent vortices \citep{burgess2017vortex} are often associated with screening \citep{jimenez2021collective,grooms2010model} and the resulting shielded vortices
often break up into tripoles \citep{carton1989generation} 
consisting of a central vortex and two satellite vortices of opposite sign 180$^\circ$ apart, as seen in both experiment \citep{van1991laboratory} 
and DNS \citep{orlandi1992numerical}. 
In fact tripolar vortices are an exact solution of the 2D Euler equation \citep{kizner2004tripole} and are known to be stable point-vortex states \citep{kizner2011stability}. 
 
We focus here on 2D turbulence driven by a parametrised force that varies continuously from purely random to pure finite-wavenumber linear instability. We show that shielded vortices \NEW{spontaneously} arise for sufficiently large instability growth rates, and that the resulting flow displays both spontaneous symmetry breaking and bistability.

\section{Setup}\label{sec:setup}
We study the 2D Navier-Stokes equation for \NEW{an incompressible} velocity field $\mathbf{u}\equiv(u,v)$,
\begin{eqnarray}
    \partial_t \mathbf{u} +\mathbf{u}\cdot \nabla \mathbf{u} &=& -\nabla p + \mathbf{f} - \nu_n (-\nabla^2)^n \mathbf{u} - \beta |\mathbf{u}|^2 \mathbf{u}, \hspace{1cm}    \nabla \cdot \mathbf{u} = 0 \label{eq:nse},
\end{eqnarray}
in the domain $D=[0,2\pi]^2$ with periodic boundary conditions, with pressure $p$, hyperviscosity $\nu_n$ of order $n$ \NEW{($n=4$ in most runs)}, damping coefficient $\beta>0$ and forcing
\begin{equation}
    \mathbf{f}=   \gamma \mathcal{L} [\mathbf{u}] + (1-\gamma)\mathbf{f}_\epsilon. \label{eq:forcing}
\end{equation}
Here $\gamma\in[0,1]$, and $\mathcal{L}[\mathbf{u}]$ is a linear operator with Fourier transform
\begin{eqnarray}
    \widehat{\mathcal{L}[\mathbf{u}]}(\mathbf{k}) = \nu_* k^2 \hat{\mathbf{u}}(\mathbf{k}),\quad \nu_*>0, \label{eq:lin_force}
\end{eqnarray}
for wavenumbers $\mathbf{k}$ in the annulus $k=|\mathbf{k}| \in [k_1,k_2]$, and $ \widehat{\mathcal{L}[\mathbf{u}] }(\mathbf{k})=0$ otherwise. This linear term is associated with a maximum growth rate $\sigma\equiv\nu_* k_2^2$ \NEW{with $\nu_*$ chosen such that the ratio $r=\gamma\sigma/(\nu_n k_2^{2n})$ between forcing and dissipation at the most strongly forced scale $k_2$ varies from $r=0$ to $r\gg 1$ as $\gamma$ increases from $0$ to $1$.} The second term in (\ref{eq:forcing}) involves the solenoidal zero-mean white stochastic force $\mathbf{f}_\epsilon(\mathbf{x},t)$ with random phases acting on a thin shell of wavenumbers centered on $k=k_2$ \NEW{(we also performed runs with $\mathbf{f}_\epsilon$ acting on all scales in $[k_1,k_2]$, and found no qualitative differences in the resulting flow)}.
Thus the mean \NEW{power injected by ${\bf f}_\epsilon$} is fixed, $\overline{\mathbf{u}\cdot \mathbf{f}_\epsilon} =\epsilon$, 
where $\overline{(\cdot)}$ is the ensemble average. Random forcing is often used in numerical studies of 2D turbulence, e.g. \cite{chan2012dynamics}, although time-independent forcing has also been used, e.g. \cite{tsang2009forced}. The choice (\ref{eq:forcing}) allows us to transition continuously from random forcing to a wavenumber-localised instability. A similar superposition of random and deterministic forcing was used by \cite{jimenez2007spontaneous} who maintained a fixed injection rate, while we consider a true instability, injecting energy at a rate proportional to the forcing-scale velocity. In a related study of an active fluid model with a negative viscosity forcing like that in eq.~(\ref{eq:lin_force}), \cite{linkmann2019phase} observed a large-scale condensate when a large viscosity is imposed on all scales below the forcing scale. This assumption results in low to moderate Reynolds numbers, a regime more amenable to study \citep{bos2020linearly}. 
Here, we employ hyperviscosity, a well-established numerical device for reducing finite-viscosity effects at moderate resolution \citep{borue1995forced}, 
and focus instead on the high-Reynolds number regime of instability-forced flow. This procedure ensures that our DNS are well resolved even for the large injection rates of energy and enstrophy at large $\gamma$. Nonlinear dissipation as in eq.~(\ref{eq:nse}) is commonly used in hydrodynamic models of Toner-Tu type \citep{toner1995long} 
and is needed here to saturate the linear instability. 

\begin{center}
\begin{table}
\centering
\begin{tabular}{|c|c|c|c|c|c|c|c|}
    \hline
    Set & $\#$ of runs & $\gamma$ & $\beta$& $n$& $Re_n $&Initial condition  \\ \hline
    A & 19 & $0-0.95$ & $1\times 10^{-4}$ &$4$& $3\times 10^{9}-4\times10^{18}$ & small-amplitude random \\
    B & 6 & $0.2-0.5$ & $1\times 10^{-4}$ &$4$& $3\times 10^{9}-5\times 10^{12}$ & large-scale condensate \\
    C & 4 & $0.9$ &$1\times 10^{-4}-5\times10^{-6}$ & $4$& $1\times 10^{18}-5\times10^{19}$  & vortex gas\\
    D & 15 & $0-0.95$ & $1\times 10^{-4}$& $1$&$5\times 10^2 - 2 \times 10^{4}$ &small-amplitude random
\end{tabular}
\caption{\NEW{Summary of runs. Sets A-C: $512^2$ resolution, hypervisocity $\nu_4=10^{-14}$, $\nu_*=0.002$ and set D: $1024^2$ resolution,  $\nu_1=0.0011$, $\nu_*=10\nu_1$. For all runs, we take $k_1=33$, $k_2=40$, giving Reynolds numbers $Re_{n} = U_{rms} L_{I}^{2n-2/3}/\nu_n$ with $U_{rms}=\sqrt{\sum_k E(k)}$ and integral scale $L_I = \sum_k 2\pi k^{-1}E(k)/ \sum_k E(k)$. Reynolds numbers are given at late times.}}
\label{tab:summary_runs}
\end{table}
\end{center}

We use the \NEW{pseudospectral} Geophysical High-Order Suite for Turbulence (GHOST) \citep{mininni2011hybrid} to perform DNS of the system (\ref{eq:nse}) \NEW{using a fourth-order Runge-Kutta scheme in time. Our runs consist} of \NEW{four} sets, summarised in Table \ref{tab:summary_runs}. Set A consists of runs with small-amplitude, random initial conditions. \NNEW{The runs in set B were initialised with a large-scale condensate obtained in set A for purely random forcing ($\gamma=0$). In set C we initialise with a vortex gas, again from set A (at $\gamma=0.9$), and vary $\beta$.}  
In all \NEW{the runs described below} we use $512^2$ resolution to be able to simulate the system for long times; \NEW{$1024^2$ runs with regular viscosity ($n=1$) were also performed \NEW{(set D)}, and are qualitatively similar to the hyperviscous runs, although they can only reach shorter times.} We record the energy $E=\langle \mathbf{u}^2 \rangle$, the enstrophy $\Omega = \left \langle \omega^2 \right \rangle $, with vorticity $\omega \equiv \partial_x v - \partial_y u$ and spatial average $\langle \cdot \rangle$, the energy spectrum
\begin{equation}
    E(k) = \sum\nolimits_{\mathbf{q}:k-\frac{1}{2}\leq|\mathbf{q}|< k+\frac{1}{2}} |\hat{\mathbf{u}}(\mathbf{q})|^2,
\end{equation}
and the spectral energy and enstrophy fluxes through wavenumber shell $k$
\begin{eqnarray}
    \Pi_E(k) =  \left \langle \mathbf{u}^<_k \cdot(\mathbf{u}\cdot\nabla \mathbf{u}) \right \rangle, \hspace{1.5cm} \Pi_\Omega(k) = \left \langle \omega^<_k (\mathbf{u}\cdot \nabla \omega) \right \rangle \label{eq:pi},
\end{eqnarray}
\NNEW{cf.~\cite{frisch1995turbulence}, Eq.~(2.52),} where $(\cdot)^<_k$ is defined as $
    f^<_k(\mathbf{x}) = \sum_{\mathbf{q}:|\mathbf{q}|\leq k} \hat{f}(\mathbf{q}) \exp(i\mathbf{q}\cdot \mathbf{x}). 
$
\vspace{-0.5cm}
\section{Overview of the results}\label{sec:overview}
\begin{figure}
    \centering
    \includegraphics[width=0.45\textwidth,height=4.2cm]{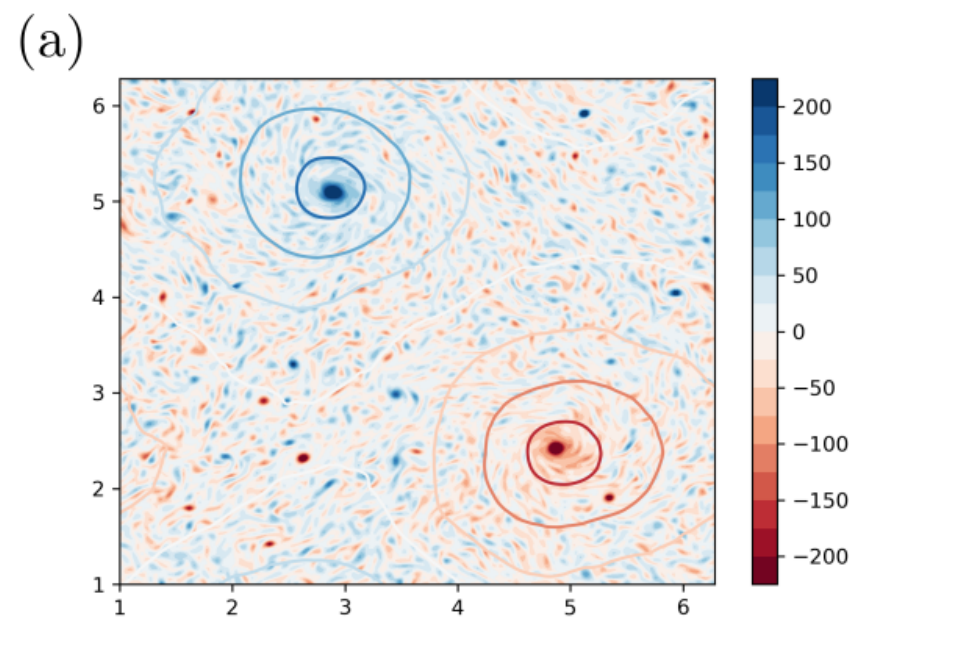}
    \includegraphics[width=0.45\textwidth,height=4.2cm]{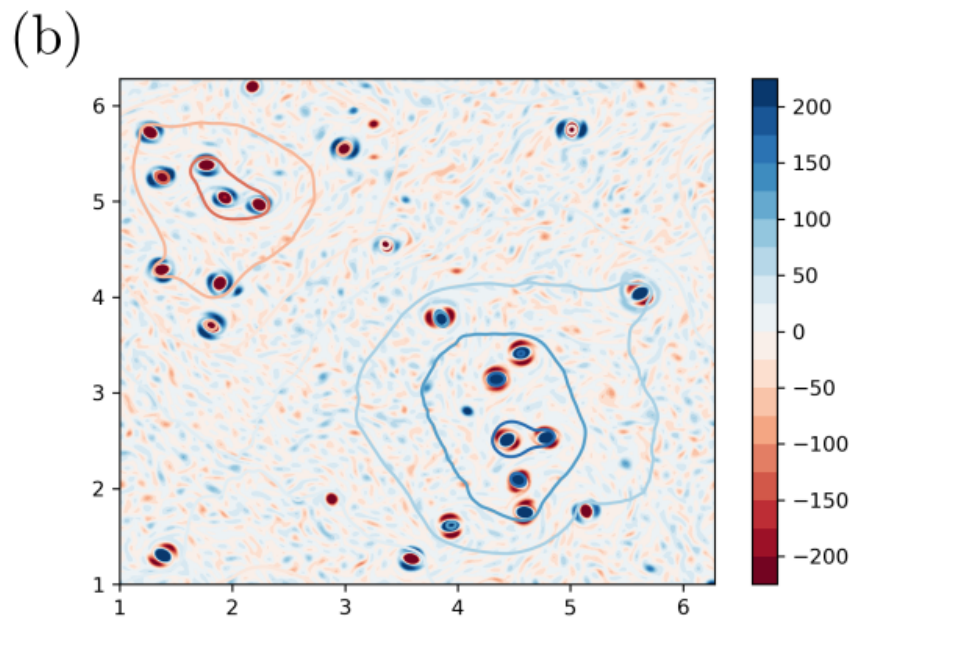}
    \includegraphics[width=0.45\textwidth,height=4.2cm]{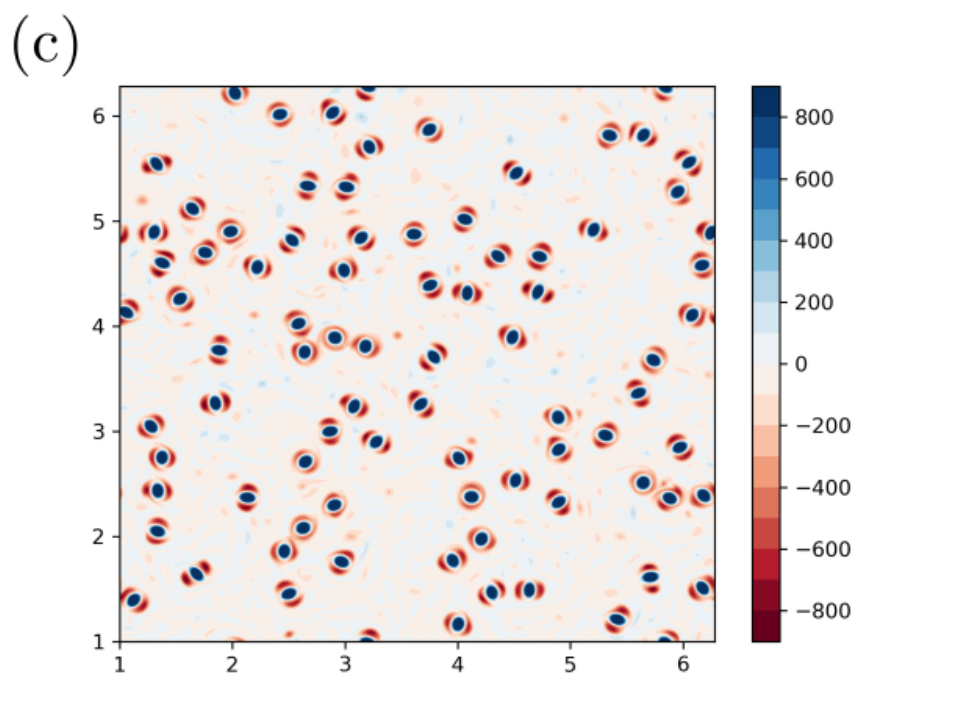}
    \caption{Visualisations of the vorticity field, with contour lines showing the streamfunction \NEW{(panels (a), (b) only)}. (a) Large-scale condensate state at $\gamma=0$. (b) Mixed state, with shielded vortices clustering in large-scale vortices at $\gamma=0.35$. (c) Dilute gas of shielded vortices at \NNEW{$\gamma=1$}. \NEW{Note the different color scale used in panel (c).}}
    \label{fig:vis}
\end{figure}

\begin{figure}
    \includegraphics[width=0.45\textwidth]{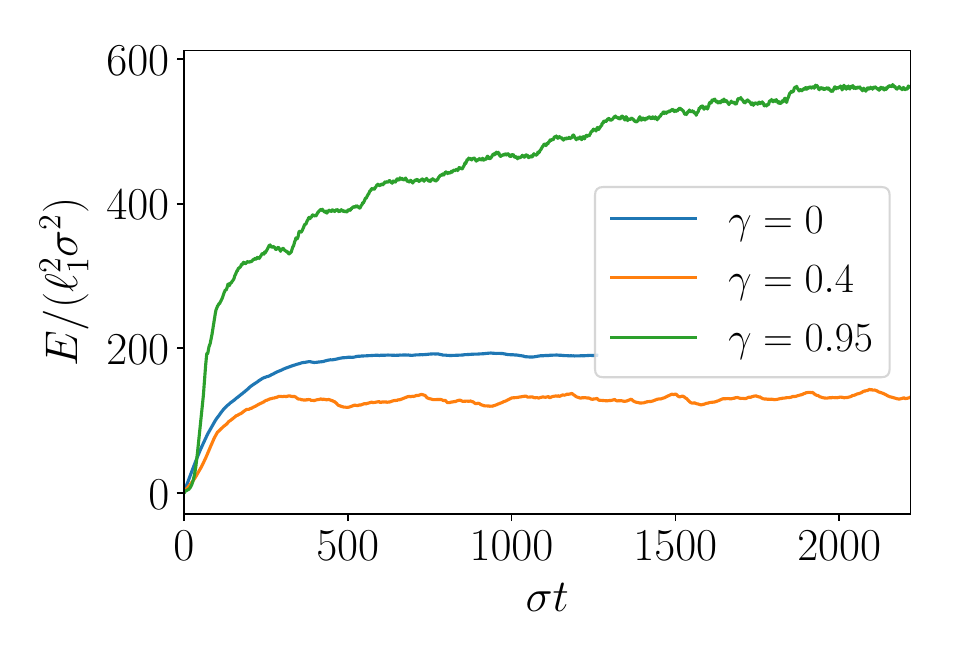}
    \includegraphics[width=0.45\textwidth]{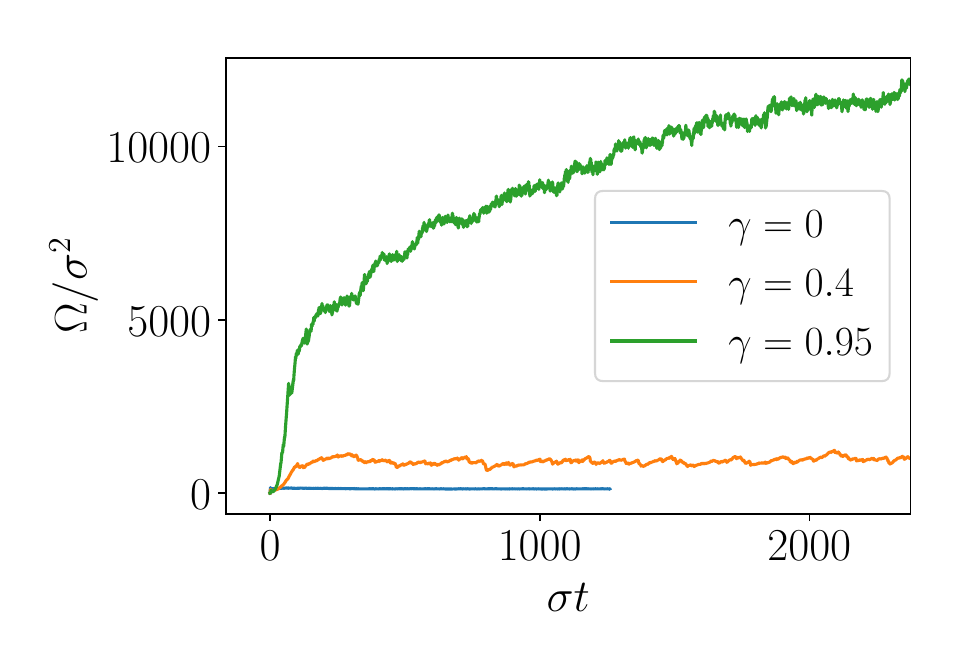}
    \caption{\NNEW{Time series of nondimensionalised energy $E$ and enstrophy $\Omega$ for a condensate state ($\gamma=0$), a mixed state ($\gamma=0.4$) and shielded vortex gas ($\gamma=0.95$). For the former two cases, a steady state is reached quickly, while in the vortex gas, energy and enstrophy grow slowly.}}
    \label{fig:ts_ens_ene}
\end{figure}

Figure \ref{fig:vis} shows \NNEW{snapshots of} typical solutions at different $\gamma$ obtained by integrating from small-amplitude random initial conditions. At $\gamma=0$ (random forcing) a large-scale condensate forms. At $\gamma=0.35$ a large-scale circulation persists, as indicated by the streamfunction, but \NEW{small tripolar vortices with positive and negative cores} appear, concentrating within the large-scale vortices of the corresponding sign. \NNEW{ We stress that the large-scale vortices evolve in time, constantly changing their position and shape.} At $\gamma=1$ a state of broken symmetry is present, with a large number of same sign vortices. We call this state a \textit{(shielded) vortex gas}. \NEW{In our runs with regular viscosity, we observed the same phenomenology as in figure \ref{fig:vis}.} \NNEW{In figure \ref{fig:ts_ens_ene}, we show the time evolution of energy and enstrophy for each of the three observed regimes. When a condensate is present, energy and enstrophy saturate quickly, but continue to grow in the case of the vortex gas. This case is discussed in much greater detail in section \ref{sec:symm_breaking}.}

\begin{figure}
    \centering
    \includegraphics[width=0.49\textwidth]{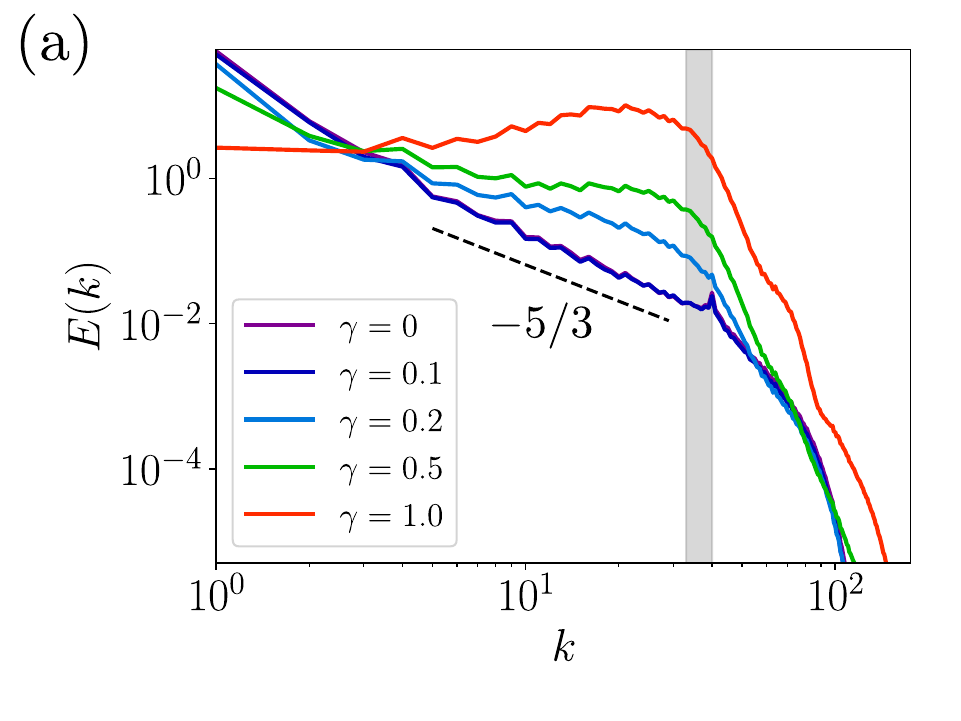} 
    \includegraphics[width=0.49\textwidth]{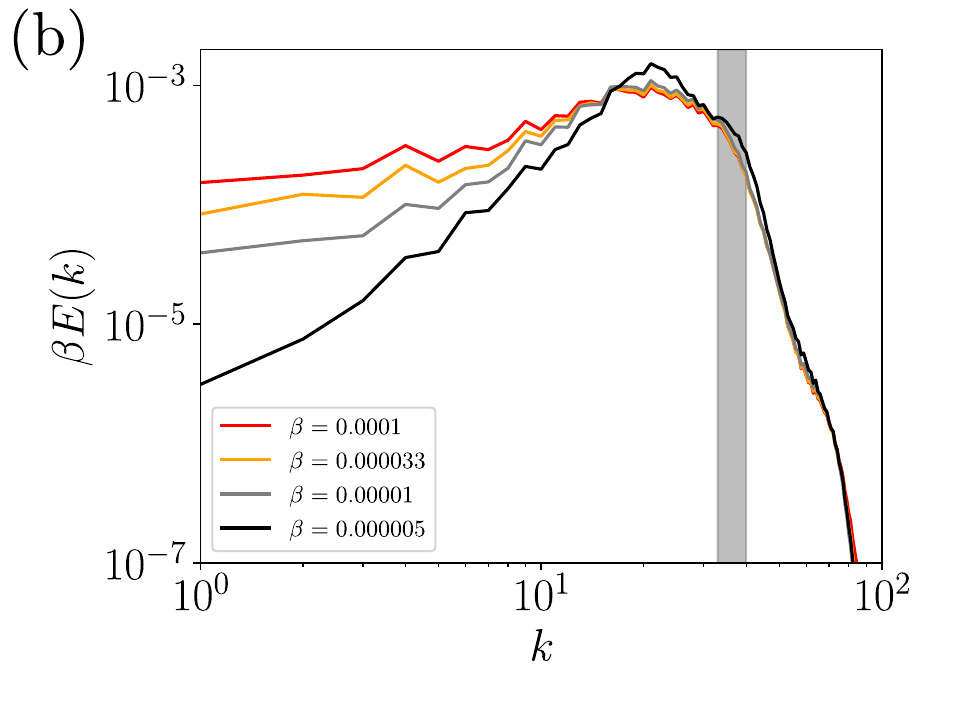}
    \caption{Energy spectra $E(k)$ at late times, averaged over the last 500 snapshots. The grey bar indicates the forcing range $[k_1, k_2]$. Dashed line indicates a $-5/3$ power law. (a) Spectra for different $\gamma$ from set A. \NEW{The same color scale used here for different $\gamma$ is used subsequently in figures \ref{fig:fluxes}, \ref{fig:vorticity_pdfs}, \ref{fig:vortex_census}.} (b) Spectra for different $\beta$ ($\gamma=0.9$, set C), scaled by $\beta$.	}
    \label{fig:energy_spectra}
\end{figure}

The spectra associated with states at different $\gamma$ are shown in figure \ref{fig:energy_spectra}(a). At $\gamma=0$ and $\gamma=0.1$, there is a \NEW{$-5/3$} power-law range at scales larger than the forcing, and a build-up of energy at $k=1$, i.e. a condensate. As $\gamma$ increases through $0.2$, the energy in the large scales decreases and that in the near-forcing scales increases, \NEW{marking} the appearance of shielded vortices. The condensate is weaker but persists at $\gamma=0.5$, while the near-forcing-scale energy grows. Finally, at $\gamma=1$ the spectrum no longer peaks at the largest scales, but rather at scales comparable to twice the forcing scale $\ell_1\equiv 2\pi/k_1$. \NEW{The spectral bumps seen at high k are likely related to harmonics of the main peak.} Panel (b) shows the spectrum in the vortex gas state \NEW{($\gamma=0.9$)} at different values of the damping coefficient $\beta$, rescaled by $\beta$. At the forcing scales and below, the rescaled curves collapse. At smaller $k$, as $\beta$ decreases, the peak near $2\ell_1$ becomes more pronounced, and the portion of energy in large scales falls, due to larger vortex amplitudes at smaller $\beta$. 

\begin{figure}
    \centering
    \includegraphics[width=0.49\textwidth]{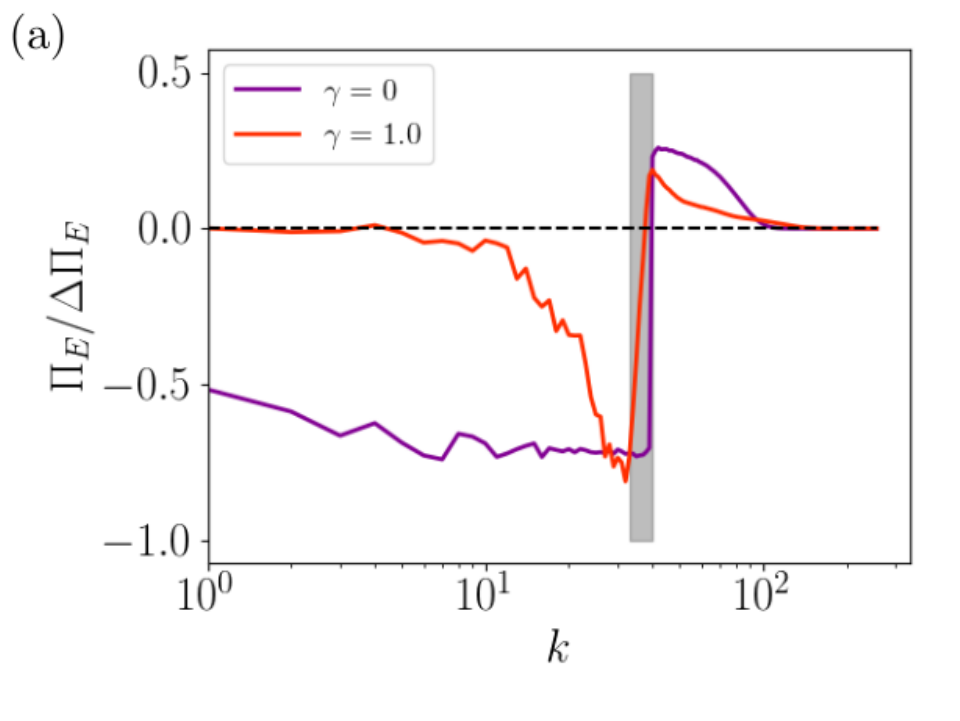}
    \includegraphics[width=0.49\textwidth]{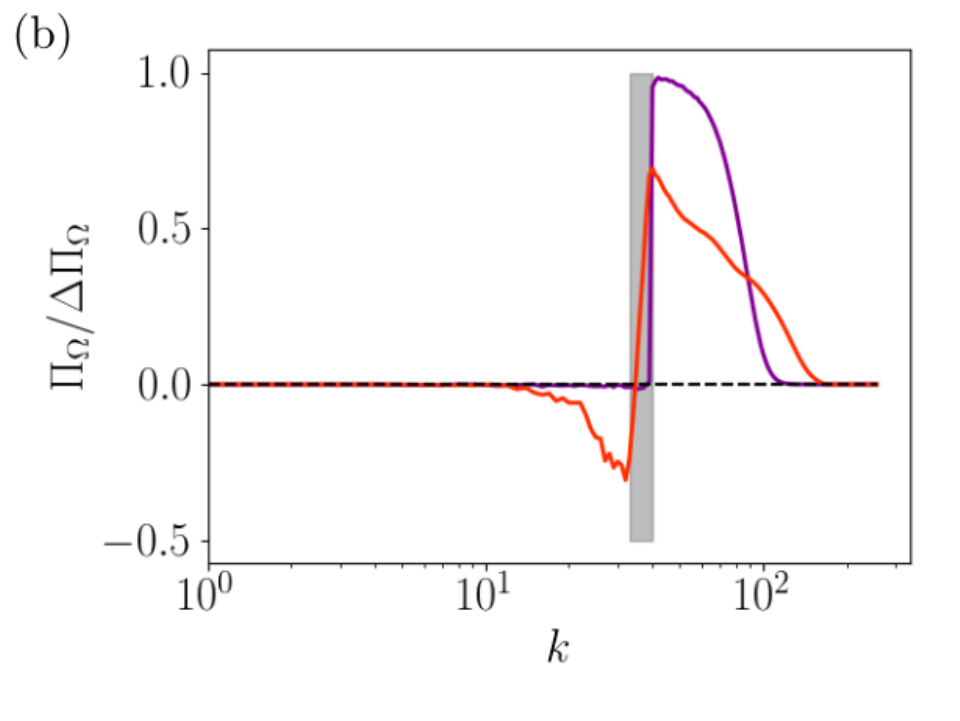}
    \caption{(a) Energy flux $\Pi_E$  and (b) enstrophy flux $\Pi_\Omega$, \NEW{rescaled by $\Delta \Pi_i = \max_{k}(\Pi_i)-\min_k(\Pi_i)$, $i=E,\Omega$}, averaged over 500 time steps at late times, for $\gamma=0$ (random forcing), and $\gamma=1$ (vortex gas).  \NEW{Unrescaled fluxes in the gas are significantly greater than at $\gamma=0$ and fluxes are strongly scale-dependent: $\Pi_E$ is suppressed at large scales, while $\Pi_\Omega$ is forward, with a weak local inverse transfer.} Shaded region indicates forcing. }
    \label{fig:fluxes}
\end{figure}

Figure \ref{fig:fluxes} shows the energy and enstrophy fluxes at $\gamma=0$ (random forcing) and $\gamma=1$ (vortex gas). The negative (i.e. inverse) energy flux in the vortex gas state is suppressed at scales larger than about $2\ell_1$. A small forward energy flux feeds the remaining finite dissipation at small scales. The forward enstrophy flux for random forcing remains so in the vortex gas although a small inverse enstrophy flux is also present, reaching to around $k_1/2$. \NEW{We stress that residual fluxes at $\gamma=1$ are strongly scale-dependent, indicating absence of self-similar cascade, in contrast to the $\gamma=0$ case where fluxes are reasonably constant.} The suppression of nonlinear transfers by coherent vortices is reminiscent of that in decaying 2D turbulence \citep{mcwilliams1984emergence}. In our case, there are two competing time scales, however: the instability growth rate at which energy is injected into the forcing scales, and the rate at which energy is transferred out of the forcing scales by nonlinearity. If energy is injected too rapidly for nonlinear transfer to remove it, it builds up near the forcing scales, and coherent vortices form, further suppressing \NEW{nonlinear fluxes}. \NEW{In the absence of any cascade mechanism, nonlinear damping is required to saturate the build-up of near-forcing-scale energy.}
\vspace{-0.5cm}

\section{Vorticity statistics and radial profiles} \label{sec:vort_stat}
\begin{figure}
    \centering
    \includegraphics[width=0.49\textwidth]{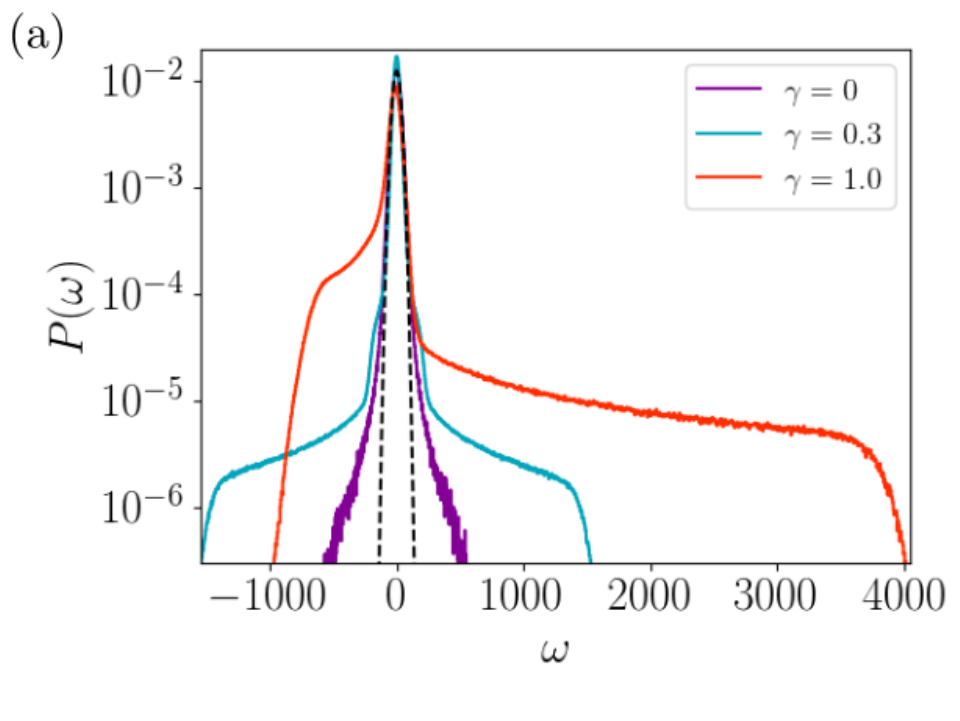}
    \includegraphics[width=0.49\textwidth]{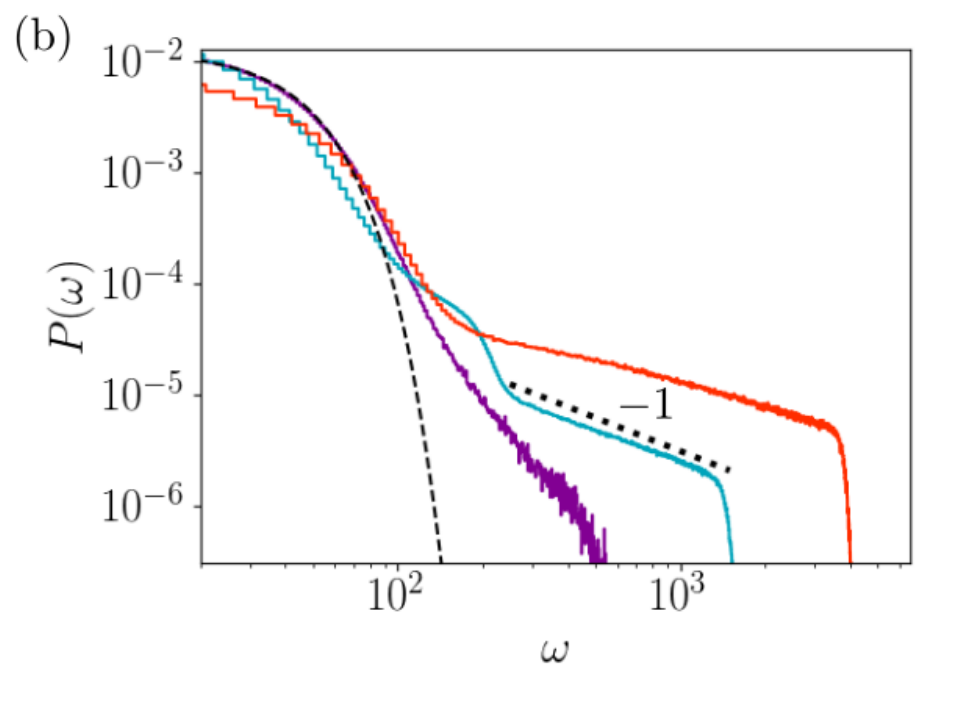}
    \caption{(a) Lin-log plot of the vorticity PDF sampled over all spatial points, aggregated over 800 snapshots, for three different values of $\gamma$ from set A. \NNEW{The dashed curve centered on $\omega=0$ is a Gaussian fit.} (b) Same quantities in a log-log plot. The tails have power-law form: the thick dashed line shows a power law with exponent $-1$.}
    \label{fig:vorticity_pdfs}
\end{figure}

\begin{figure}
    \centering
     \includegraphics[width=0.495\textwidth]{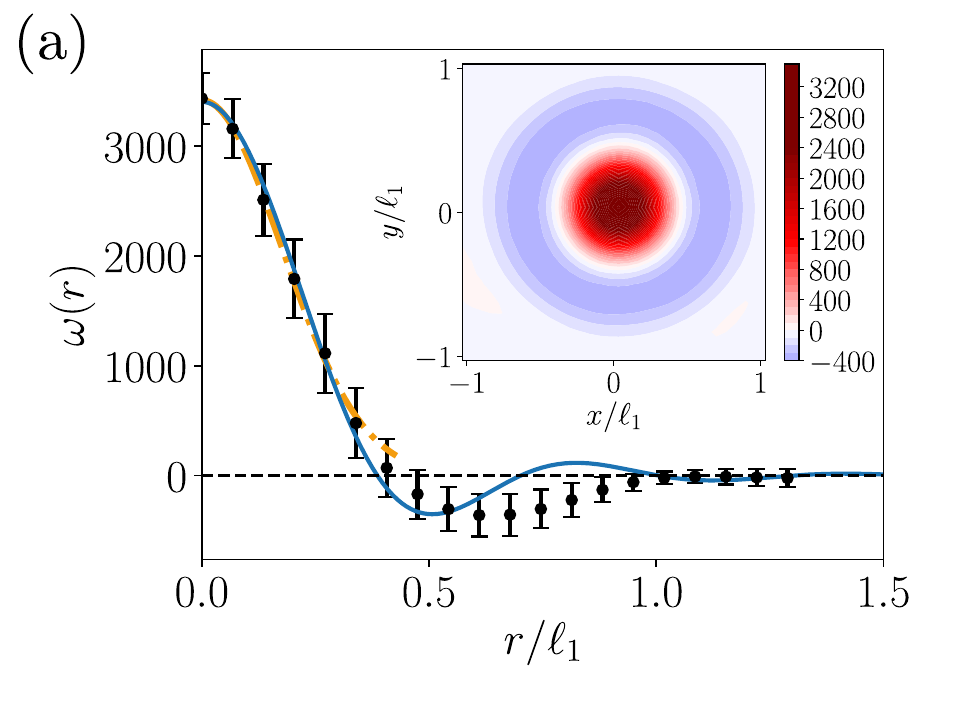}
     \includegraphics[width=0.495\textwidth]{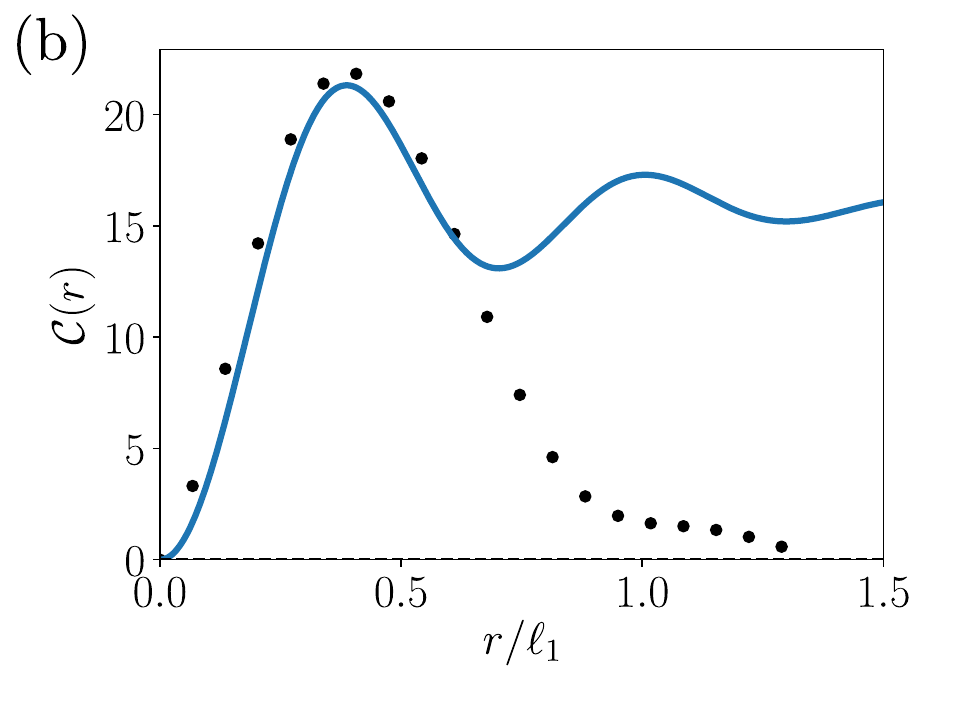}
    \caption{(a) Circles: average shielded vortex profile at $\gamma=0.9$ (set A).  \NNEW{Approximately Gaussian core (orange dash-dot line: Gaussian fit, blue solid line: hyperviscous prediction from \cite{jimenez1994hyperviscous} with hyperviscosity exponent $n=4$ as used here) surrounded by an opposite-signed shield. Radial extent is set by $\ell_1\equiv 2\pi/k_1$. Inset: population average over many tripolar vortices with arbitrary orientations. (b) Circulation $\mathcal{C}(r) = 2\pi \int_0^r \omega(r') r' dr' $ for an average shielded vortex (black circles) becomes vanishingly small beyond $r\gtrsim \ell_1$, indicating vanishing azimuthal velocity and no long-range interactions. Blue solid line: as in panel (a).}}
    \label{fig:radial}
\end{figure}

The spectra and fluxes of inviscid invariants are useful tools for analysing turbulence but have two shortcomings. First, spectral studies discard phase information and in particular vortex signs. Second, in the presence of coherent structures, such studies do not provide us with the corresponding physical space picture. To address these points, we begin by considering the statistics of the signed vorticity. Figure \ref{fig:vorticity_pdfs} shows the probability density function (PDF) of vorticity, $P(\omega)$, for three different values of $\gamma$ from set A, generated from 
over 800 snapshots. At $\gamma=0$ (random forcing) the central region of the PDF near $\omega=0$ is close to a Gaussian\NNEW{, but there are heavy tails at larger $|\omega|$.} This is consistent with the results of \cite{pasquero2002stationary}. 
For $\gamma=0.3$, the shielded vortices that are present manifest themselves in the form of \NNEW{significantly longer} heavy tails in the PDF, but the PDF remains approximately symmetric. At $\gamma=1$, the amplitude of the shielded vortices extends to larger $|\omega|$ due to stronger driving, and a pronounced skewness in the PDF develops, reflecting the broken symmetry. The log-log plot in figure \ref{fig:vorticity_pdfs} shows that the heavy tails are of power-law form with exponent close to $-1$. Power-law tailed PDFs can correspond to rare events in time or in space. Here, they represent the spatial localisation of vorticity inside shielded vortices, which are coherent over long times. Figure \ref{fig:radial}(a) shows the radial vorticity profile within the shielded vortices, computed in the vortex gas. The profile is averaged over many vortices with arbitrary orientation, resulting in an effective azimuthal average. A nearly Gaussian core is surrounded by a shield of opposite-sign vorticity, with $|\omega|=0$ at $r\approx 2\pi/k_1\equiv\ell_1$, i.e. the largest forcing scale. \NNEW{In addition to the Gaussian profile, we also compare the vorticity profile to the theoretical result of \cite{jimenez1994hyperviscous} for hyperviscous vortices, which also predicts a sign change in vorticity with radius, not unlike what we observe here. However, panel 5b shows that the circulation $\mathcal{C}$ associated with the vortices in our DNS vanishes beyond $r\geq \ell_1$. Since $\mathcal{C}=2\pi\int_0^r \omega(r) r dr = \int_{\mathcal{C}(r)} \mathbf{u}\cdot d\mathcal{\ell} $, the vanishing of $\mathcal{C}(r)$ indicates that the vortices do not generate a velocity outside this radius, and that they are thus well shielded. In contrast, the hyperviscous vortices of \cite{jimenez1994hyperviscous} are not shielded since their circulation tends to a nonzero constant as $r$ increases. For a Gaussian profile $\omega(r)$, the near circular symmetry of the core $P(\omega)d\omega \propto 2\pi r dr$ implies, in approximate agreement with figure \ref{fig:vorticity_pdfs},}
\begin{equation} 
P(\omega) \propto 2\pi r(\omega)/(\left.d\omega(r)/dr)\right|_{r=r(\omega)}  \propto \omega^{-1}.
\end{equation}
 \NNEW{We have also computed the radial vorticity profile in shielded vortices for the runs in set D, with regular viscosity (hyperviscous exponent $n=1$) and find qualitatively the same fully-shielded profiles, with the vortex size set by $\ell_1$ as in the hyperviscous runs. This indicates that the shielding is not an artifact of hyperviscosity, but rather an intrinsic result of the nonlinear dynamics.}
\section{Spontaneous symmetry breaking and vortex census} \label{sec:symm_breaking}
\begin{figure}
    \centering
    \includegraphics[width=0.075\textwidth]{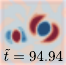}
    \includegraphics[width=0.075\textwidth]{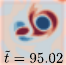}
    \includegraphics[width=0.075\textwidth]{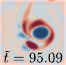}
    \includegraphics[width=0.075\textwidth]{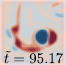}
    \includegraphics[width=0.075\textwidth]{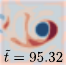}
    \includegraphics[width=0.075\textwidth]{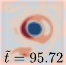}
    \includegraphics[width=0.075\textwidth]{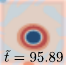}
    \includegraphics[width=0.075\textwidth]{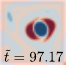} \vspace{0.4cm}\\
        \centering 
    \includegraphics[width=0.07\textwidth]{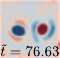}
    \includegraphics[width=0.07\textwidth]{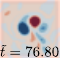}
    \includegraphics[width=0.07\textwidth]{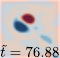}
    \includegraphics[width=0.07\textwidth]{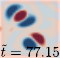}
    \includegraphics[width=0.07\textwidth]{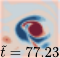}
    \includegraphics[width=0.07\textwidth]{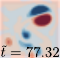}
    \includegraphics[width=0.07\textwidth]{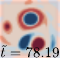}
    \includegraphics[width=0.07\textwidth]{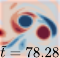}
    \includegraphics[width=0.07\textwidth]{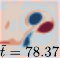}
    \includegraphics[width=0.07\textwidth]{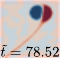}
    \includegraphics[width=0.07\textwidth]{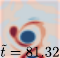}  
    \includegraphics[width=0.07\textwidth]{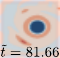}  
    \includegraphics[width=0.07\textwidth]{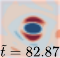}
    \caption{\NEW{Top: Absorption of a weak SV into the shield of a stronger one of opposite sign. Bottom: Life cycle of a dipole born from two colliding SVs of similar strength. 
    In both cases $\gamma=0.9$. Time is given in terms of $\tilde{t}=\sigma t$. Colors show vorticity (blue positive, red negative).} }
    \label{fig:panels1}
\end{figure}

 \noindent As mentioned, shielded vortices of one \NEW{core} sign appear in the flow at late times, provided the instability growth rate is large enough. Rotating turbulence also displays cyclone-anticyclone asymmetry \citep{bartello1994coherent} 
  but this asymmetry is the result of forced symmetry breaking.
Here, the flow maintains approximate symmetry in $\omega$ as it develops from unbiased small-amplitude initial conditions, but if $\gamma$ is large enough, this transient leads to a symmetry-broken phase where one or other sign dominates. To understand the physical space processes enabling symmetry breaking, we highlight in figure \ref{fig:panels1} two examples of typical interactions in this phase between opposite-signed shielded vortices (for $\gamma=0.9$). In the top row, a stronger vortex encounters a weaker one. The latter is 

    \begin{figure}
    \vspace{0.5cm}
    \centering 
    \includegraphics[width=0.328\textwidth]{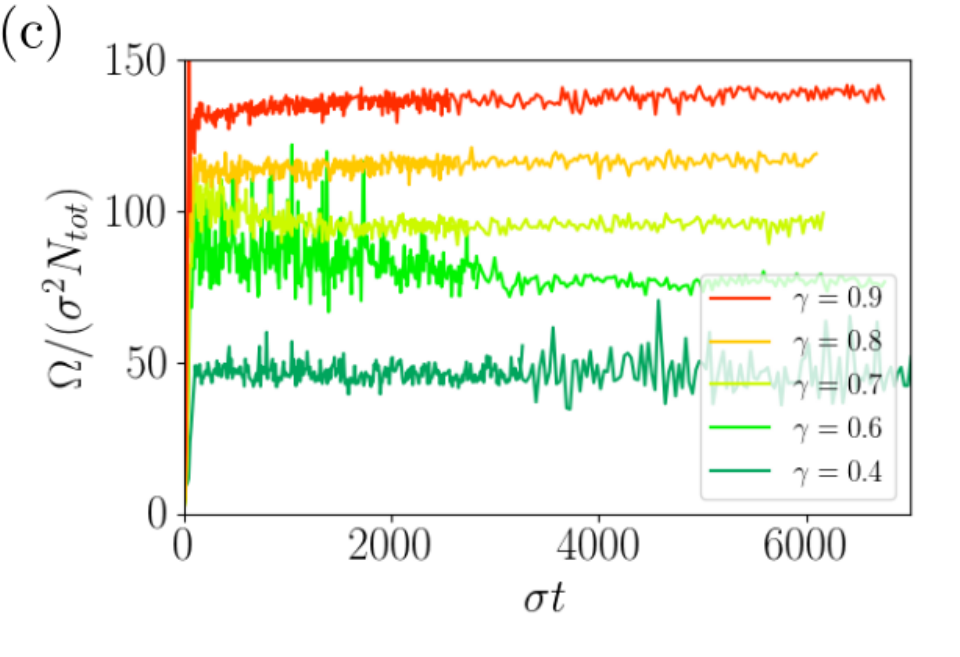}
    \includegraphics[width=0.328\textwidth]{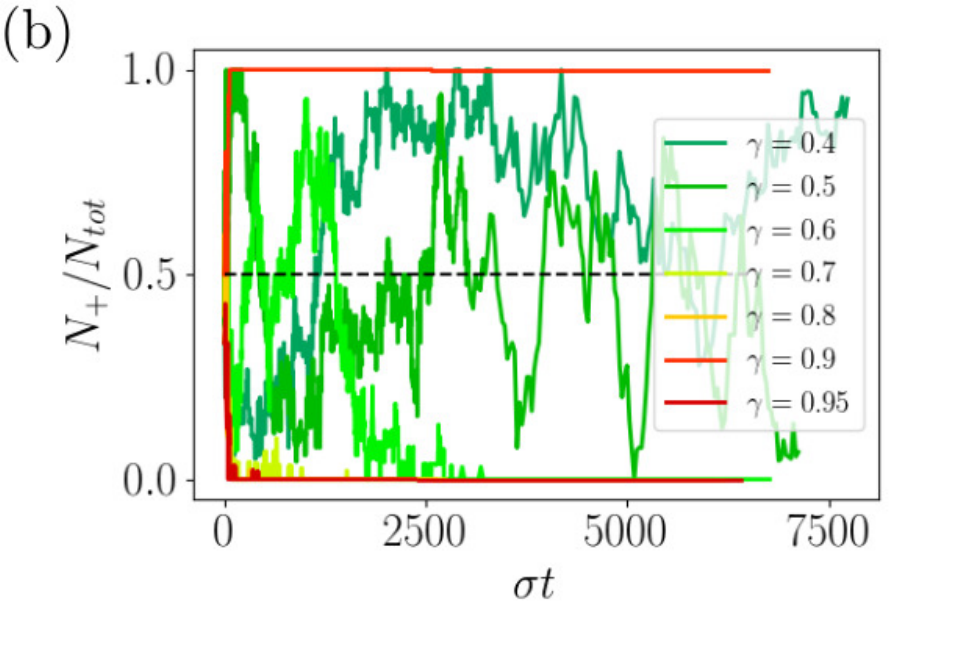}
    \includegraphics[width=0.328\textwidth]{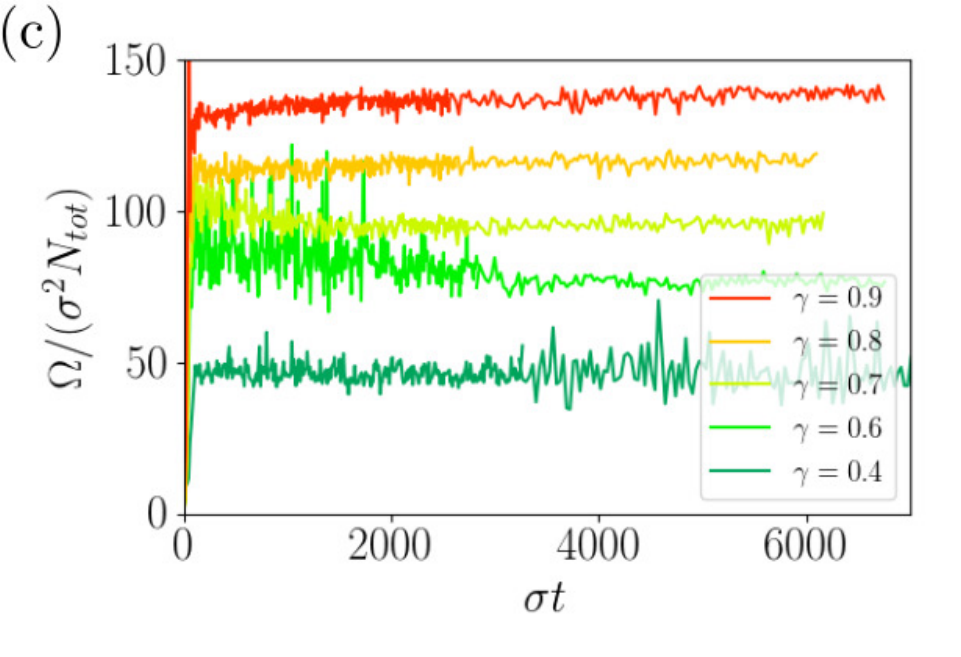}

    \caption{Vortex census data obtained as described in the text. (a) Total number of vortices $N_{tot}$ vs. time for different $\gamma$. (b) Fraction of $+$ vortices vs. time. (c) Enstrophy per vortex.}
    \label{fig:vortex_census}\end{figure}
\noindent stripped of its shield, undergoes shearing, and merges with the shield of the stronger vortex. A circular shield forms ($\tilde{t}=95.89$), which then breaks up into a tripolar one ($\tilde{t}=97.17$) as seen in experiments \citep{kloosterziel1991experimental}. The interaction described above is part of the symmetry-breaking process: from 
random small-amplitude initial conditions emerges a sea of vortices of both signs. Statistically, the populations are equal, but due to fluctuations, some vortices are stronger than their nearby opposite-signed counterparts. The interaction then eliminates the weaker vortices near stronger ones, and thus leads to a population imbalance. Asymmetric interactions are more likely at large $\gamma$, due to larger differences in the strength of vortices born at different times.

The above scenario presumes an asymmetry between interacting vortices. Figure \ref{fig:panels1} (bottom) shows a typical scenario ensuing when the vortices have comparable amplitudes: both vortices are stripped of their shields, forming a propagating dipole pair, cf.~\cite{jimenez2020dipoles}. At $\tilde{t}=77.15$, the dipole encounters a shielded vortex with a negative core, and the resulting collision strengthens the negative vortex, leading to an asymmetric dipole at $\tilde{t}=77.32$ whose curved trajectory results in a collision with a shielded vortex with a positive core at $\tilde{t}=78.19$. The result is again an asymmetric dipole, but this time with a dominant positive vortex. Note that after each collision the dominant component is determined by the core of the target vortex. In a subsequent interaction the dipole shears out the subdominant vortex into a shield around the positive core, finally becoming tripolar. The process of dipole formation, collisions and return to a single shielded vortex does not a priori favour either vortex sign. However, due to the asymmetric interaction described earlier, vortices of one sign may become more numerous. Thus, dipoles are more likely to collide with vortices of the dominant sign, thereby reinforcing their dominance.

To track the vortex population, we perform a census. For this we exploit the fact that saturated shielded vortices differ only weakly in strength, cf. figure \ref{fig:vis}. We thus filter the vorticity field, setting to zero any values with magnitude below a threshold of 75\% of the maximum vorticity magnitude (we verified that the result is insensitive to the precise choice of this threshold). Finally, 
we apply a maximum filter at the scale of vortex cores to determine the position of all vortices at a given instant. Figure \ref{fig:vortex_census} shows the total number of vortices $N_{tot}$ and the fraction of positive vortices vs. time for different $\gamma$.

Panel (a) shows that for $\gamma<0.6$, the number of vortices fluctuates around $N_{tot}\approx 20$, and does not show any increasing trend. The flow field for these runs resembles that in figure \ref{fig:vis}(b). There the positive/negative vortices are physically separated, since they \NEW{cluster} inside large-scale vortices of the same sign. For all runs at $\gamma\geq 0.6$, however, $N_{tot}$ increases with time. The larger $\gamma$, i.e. the larger the growth rate, the more rapid is this increase. Despite the long integration time of $\sigma t\lesssim O(10^4)$, $N_{tot}$ continues to slowly \NEW{drift}. The moderate resolution allows us to discern this slow trend. The increase is due to random nucleation events occurring in the sea of turbulence between vortices, which requires a vortex seed to mature without being disrupted by shear. This is likely why $N_{tot}$ rises faster at larger $\gamma$: vortices reach large amplitudes more rapidly and are harder to disrupt.   

Figure \ref{fig:vortex_census}(b) shows the fraction of vortices with positive core vs. time for different $\gamma$. At $\gamma<0.6$, this quantity fluctuates between $0$ and $1$, without converging to any particular value. At $\gamma=0.6$, a long transient, $\sigma t\sim 3000$, leads to a state of only negative vortices. At larger $\gamma$, the elimination of vortices with one sign is more rapid, but the emergent dominant sign is random. 
We quantify this transition by the enstrophy $\Omega$ defined earlier. The enstrophy per vortex \NNEW{saturates quickly, and subsequently }remains constant in time and increases with $\gamma$ (see figure \ref{fig:vortex_census}(c)). \NNEW{There is a striking separation between the fast saturation in the amplitude of individual vortices, and the slow nucleation of new vortices. }

\section{Multistability} \label{sec:multistab}
\begin{figure}
    \centering
    \includegraphics[width=1.0\textwidth]{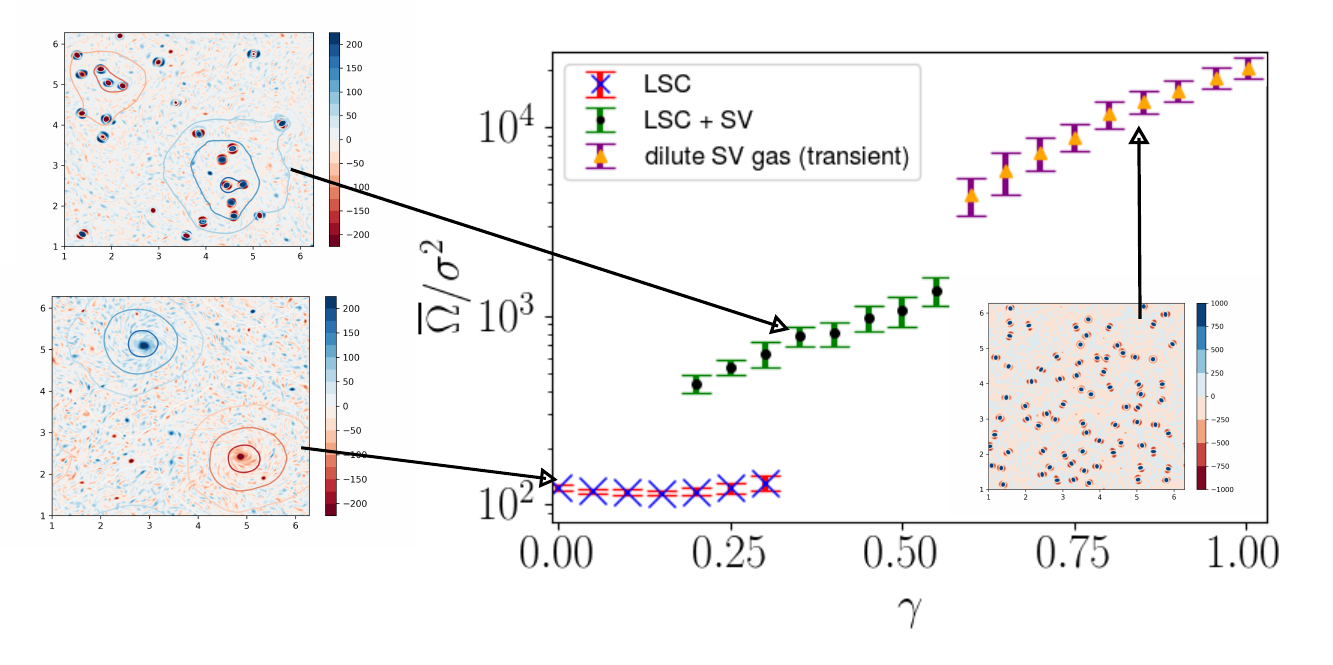}
    \caption{Summary of states at different $\gamma$ in terms of the mean enstrophy (error bars indicate standard deviation). For $\gamma\leq0.3$, a large-scale condensate (LSC) is observed. The LSC states at $0.2,0.25,0.3$ are from set $B$. Mixed states, where a LSC coexists with shielded vortices (SV) of both signs, are seen for $0.2\leq \gamma \leq 0.55$. Symmetry breaking occurs for $\gamma\geq 0.6$.}
    \label{fig:bifurcation}
\end{figure}

Figure \ref{fig:bifurcation} summarises the transitions between the different states shown in figure \ref{fig:vis}, as a function of $\gamma$, in terms of the late-time enstrophy $\overline{\Omega}$. At small $\gamma\leq 0.30$, the large-scale condensate (LSC) state exists, without any coherent shielded vortices. For $\gamma< 0.2$, the LSC states form spontaneously from small initial conditions (set A). At $0.2\leq\gamma\leq 0.3$, LSC states are stable when the flow is initialised in an LSC state (set B), but the system does not spontaneously form an LSC from small initial conditions. 
For $0.2\leq \gamma \leq 0.55$, one observes an LSC with coherent vortices of both signs, as shown in figure \ref{fig:vis}(b). Over the range $0.2\leq \gamma \leq 0.3$, there is bistability between LSC states with and without shielded vortices. For yet larger growth rates $\gamma\geq 0.6$, a state of broken symmetry forms from all initial conditions investigated. As described in the previous section, in this regime the number of vortices $N_{tot}$ steadily grows and figure \ref{fig:bifurcation} therefore shows the enstrophy at the end of each simulation. We expect that $N_{tot}$ will keep growing until a high-density, potentially crystalline, state is reached. \NNEW{A detailed study of this saturation process and of the final steady state requires very long simulations, and merits a separate study. } 

\begin{figure}
    \centering
    \includegraphics[width=0.46\textwidth]{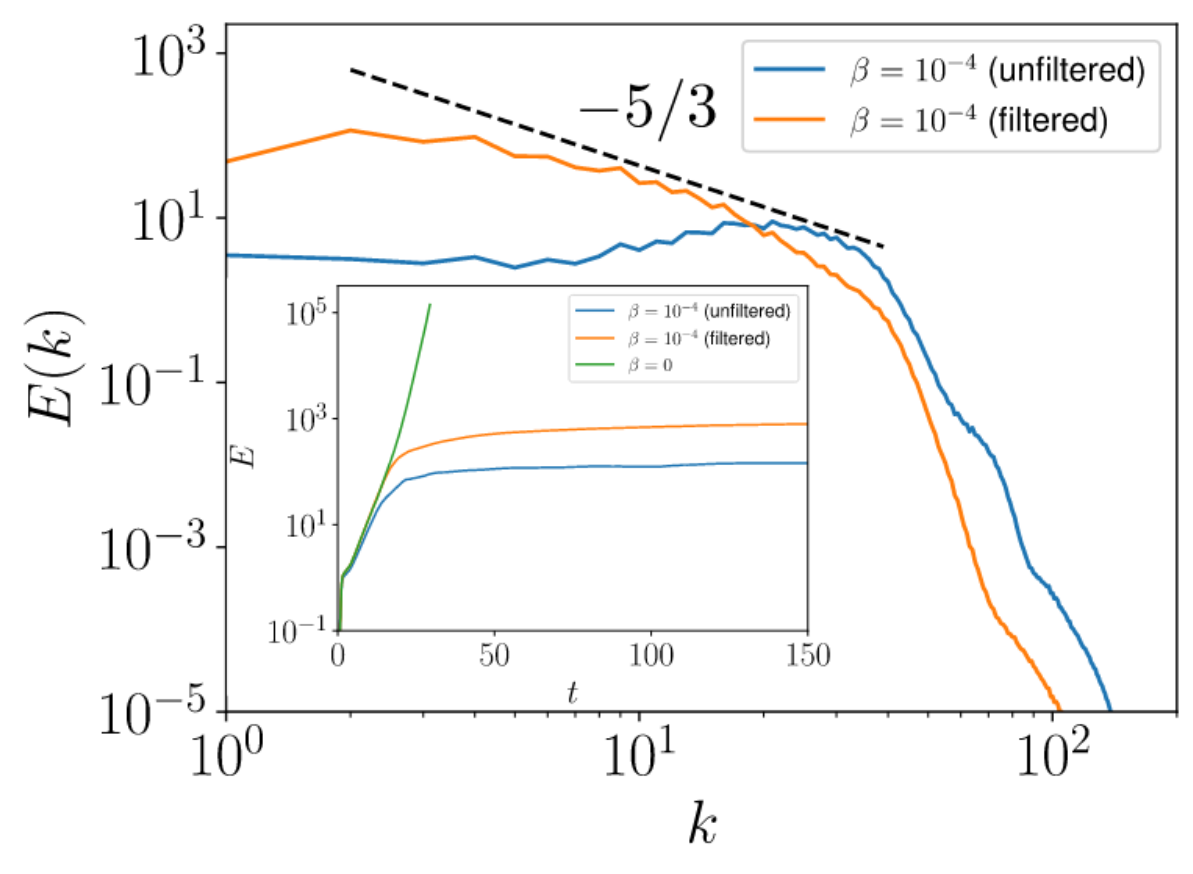}
    \includegraphics[width=0.53\textwidth]{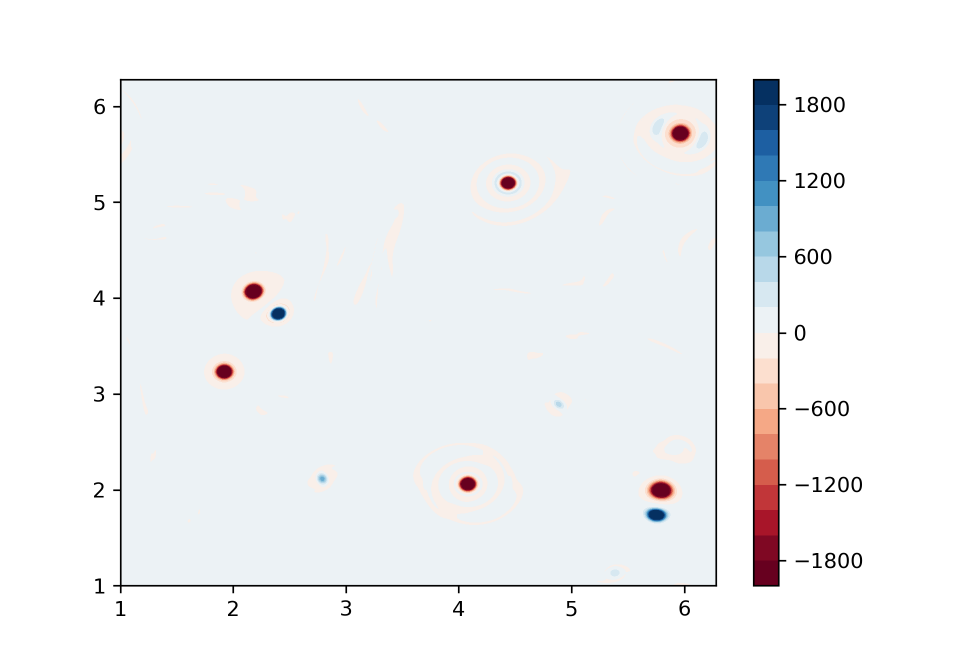}
    \caption{\NNEW{Left: Spectra for nonlinear damping $\beta |\mathbf{u}|^2 \mathbf{u}$ either full, or filtered in Fourier space to be non-zero only within the forcing range. Right: vorticity field at long time for the filtered case. Inverse energy transfer occurs when nonlinear dissipation is filtered, but is suppressed otherwise, indicating that the damping plays an important role in the suppression.}}
    \label{fig:var_diss_main}
\end{figure}

Given the rich state space shown in figure \ref{fig:bifurcation}, one may ask to what extent the results described here are specific to the choice of forcing we employed. To address this, we first tested different widths of the forcing range by varying $k_1,k_2$. \NNEW{For a given domain size, when $k_1$ is close to $k_2$, the discrete Fourier grid generates an underlying anisotropy. The dynamics in this limit are not very relevant physically, since they originate in the numerical discretisation alone. When the wavenumber shell is widened, this anisotropy disappears. In this regime, we always observe shielded vortices, provided the instability forcing is sufficiently strong.} This remains so when the dispersion relation (\ref{eq:lin_force}) is modified to a top-hat profile. In addition, we \NNEW{considered anisotropic forcing, illustrated in figure~\ref{fig:anis_forc_main}, obtained by truncating the annulus $k_1\leq |\mathbf{k}|\leq k_2$ of instability-forced wavenumbers $\mathbf{k}=(k_x,k_y)$ by requiring that $k_x$ be below some cut-off wavenumber $k_c$ \NNEW{chosen such that $k_c<k_1$,} and observed both shielded vortices and symmetry breaking for all values of $k_c$ that we  considered. We also repeated the runs in set $A$ with a modified random forcing $\mathbf{f}_\epsilon$ acting on the same scales $[k_1,k_2]$ as the instability (instead of a thin shell centered on $k_2$) and observed the same transitions as shown in figure \ref{fig:bifurcation} at approximately the same $\gamma$.} We conclude that our qualitative results are robust to changes in the details of the forcing. 
\NNEW{In addition to varying the forcing, one can also investigate the impact of nonlinear dissipation in our model. The dissipation acts on all scales, but the forcing is spectrally localised. To test whether this is relevant, we performed DNS where the nonlinear dissipation is filtered in Fourier space so it only acts on the forcing scales $[k_1,k_2]$. In this case, we find that the inverse energy cascade is no longer suppressed, as evidenced by the energy spectrum shown in figure \ref{fig:var_diss_main}. The right panel of this figure shows the corresponding vorticity field. Shielded vortices appear at early times, but the shields are subsequently lost, and do not suppress the inverse cascade. This experiment tells us that nonlinear dissipation that acts on all scales larger than the forcing scale is crucial for suppressing the inverse cascade, by dissipating energy efficiently and keeping it from reaching large scales. However, the suppression of the inverse cascade remains spontaneous, since an inverse cascade persists for random forcing, and only disappears for strong instability-type forcing. }
We mention, finally, that multistability occurs in quasi-2D turbulence \citep{van2019rare,favier2019subcritical} and beyond \citep{ravelet2004multistability} although we observed no spontaneous transitions between different branches.
\vspace{-0.4cm}
\section{Conclusions} \label{sec:conclusions}
We have shown that 2D turbulence forced by a combination of random forcing and instability differs in a fundamental way \NNEW{in the presence of damping} from the phenomenology identified by Kraichnan. Shielded vortices with Gaussian cores and forcing-scale size arise at $\gamma\approx0.2$, and undergo spontaneous symmetry breaking at $\gamma\approx0.6$. We identified interactions of opposite-signed vortices reinforcing population imbalances and enabling symmetry breaking. Bistability between condensate/mixed states occurs for $0.2\leq \gamma \leq0.3$. Such dependence of the observed flow on the forcing is an instance of non-universality, which complements other aspects of 2D turbulence that are known to be non-universal \citep{linkmann2020non}. 
\NNEW{Non-universality has also been discussed in the context of wave turbulence in the nonlinear Schr\"odinger equation \citep{vladimirova2012phase}. There, a large-scale condensate with an isotropic spectrum forms for random pumping, but for instability pumping spontaneous symmetry breaking generates an anisotropic spectrum, not unlike what we have described here.}

Although mesoscale vortices have been observed in active turbulence \citep{wensink2012meso} and 2D turbulence with hybrid forcing \citep{jimenez2007spontaneous}, we reiterate that our case differs from the former by the use of hyperviscosity (high Reynolds numbers), and from the latter by a state-dependent injection rate. \NEW{A definitive study of this system using regular viscosity and systematically varying the Reynolds will nevertheless required in the future.} It is interesting that the snapshot in figure 2 of \cite{jimenez2007spontaneous} contains a tripolar vortex, something which is hard to extract from spectral analysis without a parallel physical space perspective.


\NEW{Quasi-2D instability-driven turbulence is of particular relevance in geophysical applications. Our 2D results suggest such turbulence may behave very differently from its counterpart with \NEW{state-independent} forcing, a topic that merits further study.}


 \vspace{-0.5cm}
\section*{Acknowledgements}
\noindent \NNEW{We acknowledge the comments and suggestions from three anonymous referees, which helped improve this work. }This was work was supported by the National Science Foundation under grants DMS-2009563 (AvK, EK) and 2009319 (KJ). AvK thanks Santiago Benavides for useful discussions. \NNEW{This work used HPC resources provided by Savio (UC Berkeley), Alpine (CU Boulder) and Centre de Calcul Intensif d’Aix-Marseille, where the simulations reported here were performed.} 

\begin{figure}
        \centering
        \includegraphics[width=0.4\textwidth]{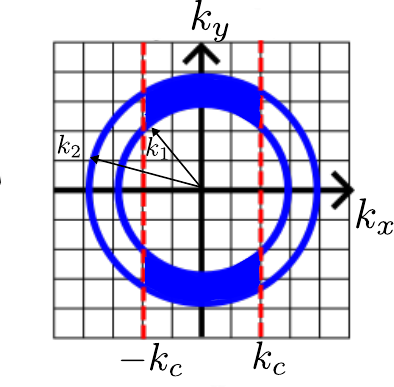}        \includegraphics[width=0.5\textwidth,height=0.40\textwidth]{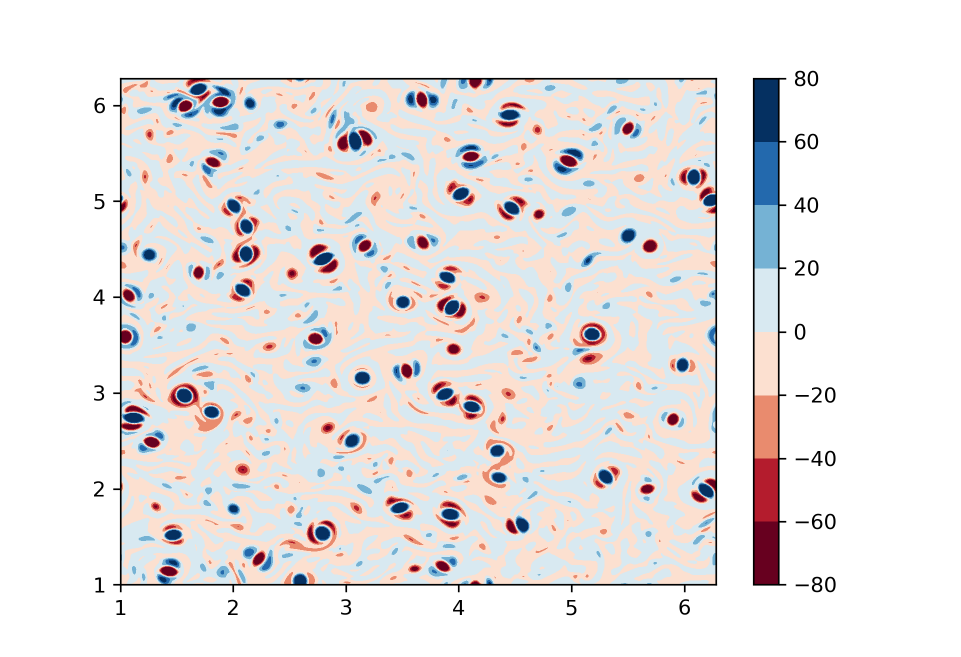}
        \caption{\NNEW{Left: illustration of the truncated wavenumber shell for anisotropic forcing. Right: vorticity field for $k_c=20$, from transient evolution, showing that shielded vortices also form for anisotropic forcing. Parameters are $k_1=33$, $k_2=40$, $\gamma=1$; remaining parameters as in set $A$.}}
        \label{fig:anis_forc_main}
    \end{figure}

\noindent \textbf{Declaration of interest:} The authors report no conflict of interest. \vspace{-1cm}
\appendix
\NNEW{
\section{Anisotropic forcing}
Here we briefly describe the case where the forcing was made anisotropic. The left panel of figure \ref{fig:anis_forc_main} illustrates how the wavenumber shell of forced modes was restricted by imposing $k_x<k_c$. The right panel shows a snapshot (from the transient regime) of the flow for $k_c=20<k_1=33<k_2=40$. Shielded vortices form, and eventually undergo symmetry breaking. The flow is qualitatively similar to that with isotropic forcing.}


\vspace{-0.5cm}
\bibliographystyle{jfm}
\bibliography{jfm_bib}

\end{document}